\documentclass[twocolumn,showpacs,preprintnumbers,a4paper,prl]{revtex4}

\usepackage{graphicx}%
\usepackage{dcolumn}
\usepackage{bm}

\makeatletter
\def\btt#1{\texttt{\@backslashchar#1}}%
\DeclareRobustCommand\bblash{\btt{\@backslashchar}}%
\makeatother

\begin{document}


\title{Peak effect in single crystal MgB$_2$ superconductor
for ${\bf H}\parallel c$-axis}

\author{M. Pissas}
\affiliation{Institute of Materials Science, NCSR,  Demokritos,
15310 Aghia Paraskevi, Athens, Greece}

\author{S. Lee, A. Yamamoto, and S. Tajima}
\affiliation{Superconductivity Research Laboratory, ISTEC, Tokyo 135-0062, Japan}%

\begin{abstract}
We have studied the phase diagram of MgB$_2$ superconductor using a single crystal
for ${\bf H}\parallel c$-axis. 
For the first time we report the existence of peak effect 
in the screening current in MgB$_{2}$ single crystal for ${\bf H}\parallel c$-axis. 
In the magnetic field regime $10<H<13.5$ kOe the local fundamental diamagnetic 
moment displays a very narrow diamagnetic step, with a temperature width of the same size as the 
zero dc-magnetic field transition. For higher field this step is transformed to a peak which 
is related with the peak effect in the screening current. Finally, for $H<10$ kOe the diamagnetic step is transformed to
a gradual transition. Our findings for the vortex matter phase diagram for the MgB$_2$
are closely related with theoretical predictions concerning the vortex matter phase diagram
of a type II superconductor in the presence of weak point disorder.
\end{abstract}
\pacs{74.60.Ge, 74.60.Jg,74.60.-w,74.62.Bf}
\maketitle


The recent discovery\cite{nagamatsu01} that MgB$_2$ compound is a superconductor with 
remarkably high transition temperature $T_{c}\sim 39$ K has generated extensive scientific
research (for a review see Ref. \onlinecite{buzea01}). 
MgB$_2$ is an anisotropic \cite{lima01,lee01,xu01,simon01,budko01,Papav}
type II superconductor with a sample dependent anisotropic constant 
$\gamma=H_{\rm c2}^{\rm ab}/H_{\rm c2}^c$, 
taking values
\cite{sologubenko02a,sologubenko02b,sologubenko02c,eltsev02,pradhan01,manzano02,angst02}
into interval $2\leq\gamma \leq 6$.
The absence of weak link problem \cite{bulk} in polycrystalline MgB$_2$ samples
gives expectations for its using in practical applications.
Furthermore, the intermediate $T_c$ of MgB$_2$, in comparison
with high-$T_c$ and low-$T_c$ conventional superconductors,
makes significant the experimental exploration of its vortex matter phase diagram,
considering MgB$_2$ as a model physical system where both thermal fluctuations 
and disorder must be treated on an equal footing.
In addition, the MgB$_2$ gives us the opportunity for experimental verification of
theoretically proposed issues, like the  melting transition of the vortex matter, 
the Bragg and amorphous vortex phases 
\cite{review}.

In this  paper, we report a detailed study of the vortex matter phase diagram for 
${\bf H}\parallel c$-axis.
In the magnetic field regime $10<H<13.5$ kOe the local fundamental diamagnetic 
moment displays a very narrow diamagnetic step, with a temperature width of the same size as the 
zero dc-magnetic field superconducting transition. For higher fields this step is transformed 
to peak which is related with the peak effect. 
Finally, for $H<10$ kOe the diamagnetic step is transformed to
a gradual transition. Our findings for the vortex matter phase diagram for the MgB$_2$
are closely related with the theoretical ideas proposed recently 
\cite{giamarchi,grtas96,gingras96,koshelev98,vinokur98,kierfield98,mikitik01,radzyner02}.

\begin{figure}[htbp]\centering
\includegraphics[angle=0,width=\columnwidth]{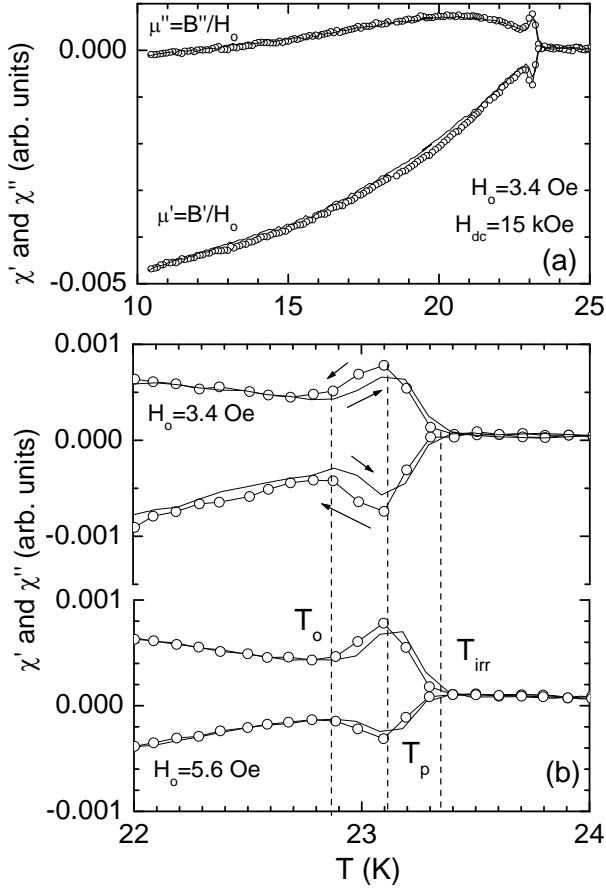}
\caption{
(a) Temperature variation of the real $\chi'$ and imaginary $\chi''$ local fundamental susceptibility
for  $H_{\rm o}=3.4$ Oe ($H_{\rm dc}=15$ kOe).
(b) Zoom of the measurements near the diamagnetic onset (for  $H_{\rm o}=3.4, 5.6$ Oe and
$H_{\rm dc}=15$ kOe) where the hysteretic behavior is observed.
}
\label{fig1}
\end{figure}
MgB$_{2}$ single crystals have been grown under high pressure in the  
quasiternary Mg-MgB$_2$-BN system at a pressure of 
4-6 GPa and temperature 1400-1700$^{\rm o}$C for 5-60 mins, in a BN container, using 
a cubic-anvil press (TRY Engineering). The present experiments were performed on
a small ($250\times 250\times 40$ $\mu$m$^3$ ) MgB$_2$ single crystal using as magnetic induction sensor a GaAsIn
Hall sensor with an active area of $50\times 50$ $\mu $m$^2$, superimposing ac ($H_{\rm ac}=H_{\rm o}\sin (2\pi ft),$
$f=0.8$ Hz) and dc magnetic fields parallel to the crystal's $c$-axis
(${\bf H}_{{\rm dc}}\parallel {\bf H}_{{\rm ac}}\parallel c$).
The real and imaginary part ($V=V'+iV''$) of the modulated Hall voltage, 
which is proportional to the local magnetic induction ($V\propto B_z$), in the surface of the crystal, was
measured by means of two lockin amplifiers. 
Measurements were performed as a function of temperature
(isofield measurements) and also as a function of the applied
field (isothermal measurements). As cryogenic environment and 
for the dc-field production  a 10 T OXFORD cryostat has been used.
Besides the superior sample quality, one of the most important aspects 
of our measurements is the sample's microscopic size, restricting to
a minimum any residual inhomogeneities.

Measurements of local magnetic induction in zero dc-magnetic field for $H_{\rm o}=1.4$ Oe,
display a $T_c=38.3$ K with a transition width (defined at the levels 10\% and 90\% of
the real part of $B'$ ) $\Delta T\approx 0.16$ K, indicating a high quality single crystal. 
Fig. \ref{fig1}(a) shows the real and imaginary parts of the local fundamental 
susceptibility ($\chi'=V'/H_{\rm o}-1$, $\chi''=V''/H_{\rm o}-1$)
as function of temperature, 
measured under a dc magnetic field of $H_{\rm dc}=15$ kOe for ac-field $H_{\rm o}=3.4$ Oe. 
The measurements have been taken during cooling and heating. As temperature decreases both $\chi'$ and $\chi''$ traces
remain zero.
Right below a temperature, which is denoted by $T_{\rm irr}$, both real and imaginary 
parts form a peak, with a peak-width roughly 0.5 K. 
Moreover, the characteristic points of the peak,
the onset temperature $T_{\rm o}$, the location of the peak, $T_{\rm p}$, and $T_{\rm irr}$ do not depend 
on the amplitude of the ac-field. The same holds and for the other dc-fields where the peak effect is present (vide infra).
As temperature further decreases both $\chi'$ and $\chi''$ exhibit the characteristic functional form of a superconducting 
sample, which supports a screening current, increasing monotonically down to $T=0$. 
The broad second maximum in $\chi''(T)$-curve, implies a small critical current density (low pinning).
More importantly, Fig. \ref{fig1}(b) shows that in the region $T_{\rm o}<T<T_{\rm p}$, 
the curves measured during cooling are above those measured during heating. 
Our observation can be attributed to the so-called peak effect which appears both in 
low and high $T_c$ superconductors. 
To the best of our knowledge it is the first time that such an experimental result concerning
the MgB$_2$ is reported.
At the $T_{\rm irr}$ the vortex lattice, due to the pinning of flux lines,
can support a finite critical current which leads to the screening of ac magnetic field.
Both the temperature and field dependance of screening current display a peak.
Different scenarios have been suggested for the peak effect. The explanation for the peak effect in terms of flux 
lattice properties was first suggested  by Pippard\cite{pippard69}.
Pippard's idea was that the energy to shear a flux lattice elastically goes to zero near $H_{c2}$
more rapidly than the pinning energy. This allows the lattice to become more
distorted near $H_{c2}$. It adjusts to increase pinning energy and thus has a higher critical 
current.
Subsequently, Larkin and Ovchinnikov\cite{larkin79} interpreted the peak effect 
based on the hypothesis that the elastic moduli of the vortex lattice suddenly soft while
going from local to nonlocal elasticity. Recently the onset of the peak effect has been associated 
with the proliferation of dislocations in the flux-line lattice
\cite{giamarchi,grtas96,gingras96,koshelev98,vinokur98,kierfield98}.
\begin{figure}[htbp]\centering
\includegraphics[angle=0,width=\columnwidth]{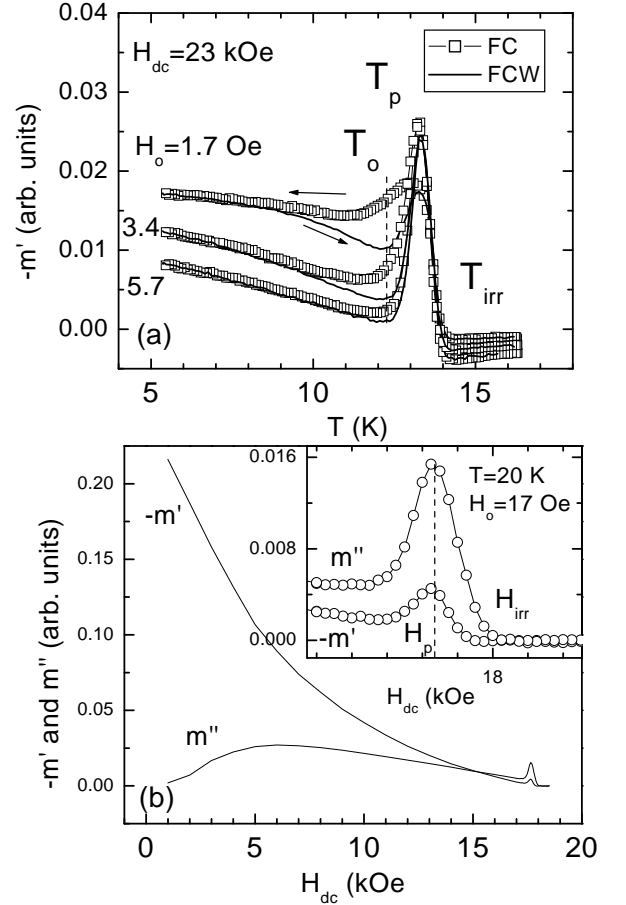}
\caption{
(a) Real $-m'=H_{\rm o}-V'$ local fundamental ac-magnetic moment
as a function of temperature, for  $H_{\rm o}=1.7, 3.4$ and 5.7 Oe
($H_{\rm dc}=23$ kOe). The data with open squares (solid line) have been measured 
during cooling (warming after field cooling), respectively.
(b) (a) Real $m'=V'-H_{\rm o}$ and imaginary $m''=V''-H_{\rm o}$ 
local fundamental ac-magnetic moment
as a function of dc-magnetic field at $T=20$ K for $H_{\rm o}=17$ Oe.
The inset shows the detail of the measurement in the region of the peak effect.
} \label{fig2}
\end{figure}

Fig. \ref{fig2} shows the real part of the local magnetic moment near the peak regime 
for $H_{\rm dc}=23$ kOe, for various ac fields, during cooling and heating.
In the particular magnetic field for $H_{\rm o}=1.7$ Oe the measurement shows a strong hysteretic
behavior for  $T<T_{\rm p}$. This hysteretic behavior reduces as the amplitude of ac-field 
increases\cite{stamopoulos02}.
Indeed, in the measurement for $H_{\rm o}=5.7$ Oe the hysteretic behavior is negligible. 
In addition, the width of the peak now is $\sim 1$ K in comparison with the width observed 
for $H_{\rm dc}=15$ kOe.
More importantly, the peak is more obvious as the amplitude of ac-field increases.
The peak effect has also been observed in isothermal measurements, as Fig. \ref{fig2}(b)
illustrates. 
The observed thermomagnetic history dependence of the ac-response is not
compatible with the conventional critical-state model. This model treats the
critical current, $J_c$ as a single valued function of the magnetic induction $B$
and temperature $T$, while our measurements indicate that $J_c$ depends on the 
measuring path in the regime $T<T_{\rm p}$. 
The observed behavior can be understood as follows: as
we expose the system to an  applied dc-field followed by cooling, through
the $H_{c2}^{c}$-line, the topological defects remain as temperature decreases
(the field cooling disordered phase is simply supercooled from the phase existing
 above $T_{\rm p}$).
In other words, the field cooling state yields a disordered vortex glass phase.
During heating this disordered state becomes more order with consequent lower
critical current in comparison with the field cooling one.
The observation of the peak effect, with negligible thermomagnetic history effects,
for large amplitude of the ac-field, may be related with the 
unblocking of the vortices from their pinned meta-stable configuration.
The ac-field triggers a transition into the stable low pinning state which does not change on
subsequent larger ac-field measurements.

\begin{figure}[htbp]\centering
\includegraphics[angle=0,width=\columnwidth]{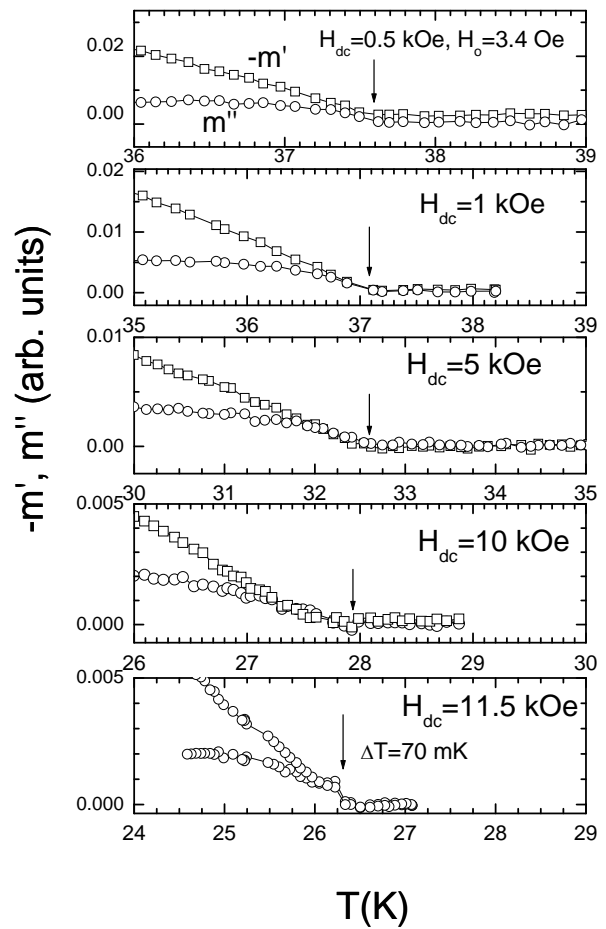}
\caption{Real $-m'=H_{\rm o}-V'$, and imaginary, $m''=V''-H_{\rm o}$ 
local fundamental ac magnetic moment
as a function of temperature, for  $H_{\rm o}=3.4$ Oe
($H_{\rm dc}=0.5,1, 5, 10$ and 11.5 kOe). } \label{fig3}
\end{figure}

More interestingly, detailed measurements in the regime $8\leq H< 15$ kOe revealed that the 
peak effect has been transformed to a very narrow diamagnetic step with a temperature 
width of similar size (or less) as the zero dc-field transition. 
Moreover, for fields $H_{\rm dc}\leq 10$ kOe (see Fig. \ref{fig3}) 
the diamagnetic step at the onset of the transition becomes 
a gradual transition (see Fig.\ref{fig3}). 
Our findings concerning the diamagnetic step in $m'$ are in accordance with the magnetoresistance
results of Eltsev et al. \cite{eltsev02} where an extremely sharp drop in the magnetoresistance
has been observed, in the same magnetic field regime.
Our observations resemble the  vortex flux lines lattice melting transition
observed in YBa$_2$Cu$_3$O$_{7-\delta}$ \cite{crabtree99} near critical points.
Based on this analogy is plausible that in this regime the transition concerns
a first order transition of the vortex lattice. Below and above the low ($H_{\rm lcp}$) and
 upper ($H_{\rm ucp}$) boundaries of this regime
the transition is transformed to a second order.

\begin{figure}[htbp]\centering
\includegraphics[angle=0,width=\columnwidth]{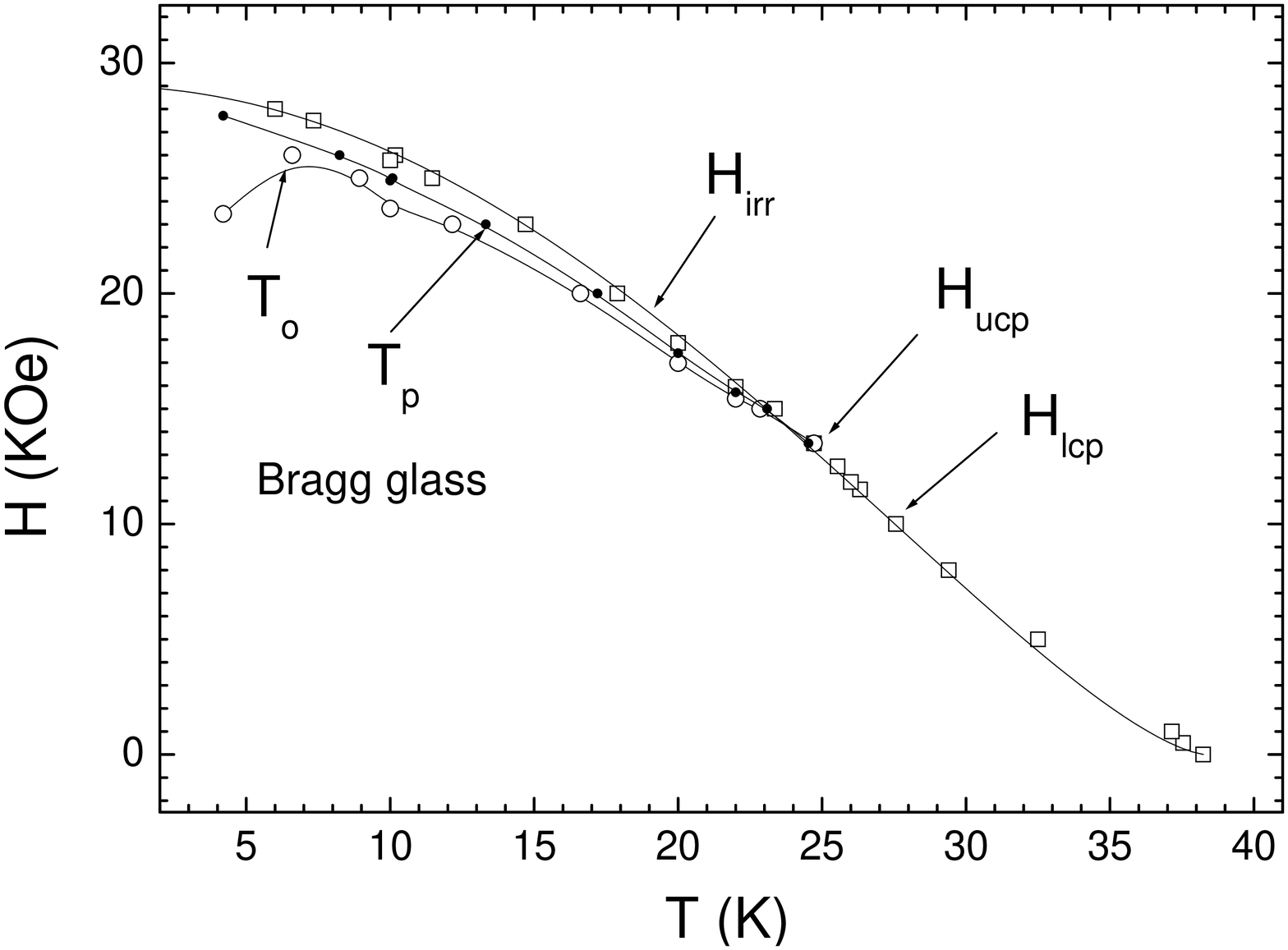}
\caption{The phase diagram of vortex matter of MgB$_{2}$ compound
for ${\bf H}\parallel c$-axis.
Presented are the onset point, $T_{\rm o}$, 
(open circles), the peak temperature,  $T_{\rm p}$, (solid circles),  and the
irreversibility points, $T_{\rm irr}$, (open squares). The 
solid line through ($T,H_{\rm irr}$)-points is a plot of the  
empirical formula $H_{\rm irr}(T)= 29[1-(T/T_c)^2]^{1.45}$.
The $H_{\rm ucp}$ and $H_{\rm lcp}$ are the points where the peak effect and the 
diamagnetic step disappear respectively.
} \label{fig4}
\end{figure}

Based on the local measurements a phase diagram for MgB$_2$ has been constructed. 
In Fig. \ref{fig4}
plotted are the onset points of the appearance of diamagnetic local moment 
that coincide with the end point of peak effect.
If the pinning of the vortex lattice starts exactly below the upper critical field these
points correspond to the upper critical field $H_{c2}^{c}$ of MgB$_2$. 
In transport measurements the $H_{c2}$ line is determined by the onset of the
magneto-resistance drop \cite{eltsev02}. 
The locus of these points is distinctly different from our $H_{\rm irr}$, but most probably the onset
of non-ohmic behavior\cite{0203337} has to do with surface superconductivity.


As illustrated in Fig. \ref{fig4}, for $T>T_c/2$, the  $H_{\rm irr}$ line displays 
a positive curvature, 
while for $T<T_c/2$ it displays a negative curvature and approaches zero temperature with
nearly zero slope at $H_{\rm irr}(0)\simeq 29$ kOe. 
The experimental points can be reproduced very well
using the empirical formula $H_{\rm irr}(T)= 29[1-(T/T_c)^2]^{1.45}$.
Included also are the points where the peak is located, as well as
the onset. The region at which the peak effect occurs, occupies a small fraction of the region 
of the mixed state, located slightly below the $H_{\rm irr}(T)$ line. 
In the regime between 10 and 15 kOe the peak effect is transformed to a very narrow
diamagnetic step and finally, for lower fields the transition becomes gradual.

It is widely accepted that the vortex phase diagram in the presence of 
weak point disorder consists of three generic phases, the vortex liquid, the high field 
amorphous vortex glass and the low field, low temperature Bragg glass \cite{review}.
These phases are governed by the three basic energies: the energy of thermal fluctuations,
pinning and elastic energies. The transition lines are determined by matching any
of the two basic energies and the match of all three energies marks the tricritical point 
\cite{giamarchi,grtas96,gingras96,koshelev98,vinokur98,kierfield98}.
Based on this unified picture we associate the line defined by the onset of peak effect to the
transition of a Bragg glass phase to an amorphous vortex glass. The point where the peak effect
is terminated can be attributed to an upper tricritical point. At this point the Bragg glass 
transition line intersects the $H_{\rm irr}$-line \cite{mikitik01}. 
Based on the sharpness of the onset of the
diamagnetic local moment in the regime $10<H<13.5$ kOe, the transition in this regime resembles
the first order melting transition. Finally, at the regime $0<H<10$ kOe the transition
becomes second order.
Since the thermal energies ($\sim kT$) and the fluctuation 
effects ($\gamma\lambda^2/\xi$) for MgB$_2$ are of the same order of magnitude
as the high-$T_c$ superconductors this scenario is very realistic.   

In the case of Nb there is structural evidence \cite{ling01} for a first-order vortex 
solid-liquid transition
at the peak temperature of the peak effect. In analogy, MgB$_2$ may exhibit a melting 
of the vortex lattice at the peak temperature, $T_{\rm p}$.
Above $T_{\rm p}$ we have a vortex liquid
up to $T_{c2}$ where the sample transits from the mixed state to the normal.

In summary, we experimentally estimated the vortex matter phase diagram  for MgB$_2$
 single crystal
superconductor for ${\bf H}\parallel c$-axis. We found that in a narrow temperature
(or field) regime a peak in the screening current exists.
The three distinct behaviors of the onset of the diamagnetic local moment resemble  
the picture of the melting transition with two critical points.

\begin{acknowledgments}
This work was partially supported by the New Energy and Industrial Technology Development 
Organization (NEDO) as collaborative research and development of fundamental technologies 
for superconductivity applications
\end{acknowledgments}


\begin{references}

\bibitem{nagamatsu01}  J. Nagamatsu {\it et al.}, Nature(London) {\bf 410}, 
63 (2001). 
\bibitem{buzea01} C. Buzea and T. Yamashita, Supercond. Sci. Technol. {\bf 14}, 
R115 (2001).

\bibitem{lima01}  O. F. de Lima {\it et al.}, \prl {\bf 86},5974 (2001); 
\bibitem{lee01} S. Lee {\it et al.,} J. Phys. Soc. Jpn {\bf 70}, 2255 (2001);
\bibitem{xu01} M. Xu {\it et al.,} \apl {\bf 79},2779 (2001).

\bibitem{simon01}  F. Simon {\it et al.}, \prl {\bf 87}, 047002 (2001).

\bibitem{budko01} S. L. Bud'ko, {\it et al.}, \prb {\bf 64}, 180506 (2001);
S. L. Bud'ko, {\it et al.}, cond-mat/0201085.

\bibitem{Papav} G. Papavassiliou et al.,
\prb {\bf 65}, 012510 (2002).


\bibitem{sologubenko02a} A.V. Sologubenko et al.,
\prb {\bf 65}, 180505R (2002).
\bibitem{sologubenko02b} A. V. Sologubenko et al.,
cond-mat/0201517.

\bibitem{sologubenko02c} A.V. Sologubenko et al.,
cond-mat/0112191.

\bibitem{eltsev02} Yu Eltsev et al.,
\prb {\bf 65}, 140501(R) (2002).

\bibitem{pradhan01} A. K. Pradhan et al.,
\prb {\bf 64}, 212509 (2001).

\bibitem{manzano02} F. Manzano et al.,
\prl {\bf 88}, 047002 (2002).

\bibitem{angst02} M. Angst et al.,
\prl {\bf 88}, 167004 (2002).

\bibitem{bulk}
S. L. Bud'ko et al.,
Phys. Rev. B {\bf 63}, 220503R (2001);
Y. Bugoslavsky et al. 
London, {\bf 411},561, (2001);
M. Pissas, et al.,
J. of Superconductivity {\bf 14}, 615 (2001).

\bibitem{review}  G. Blatter et al., Rev. Mod. Phys. {\bf 66}, 1125 (1994);
E. H. Brandt, Rep. Prog. Phys. {\bf 58,} 1465 (1995);T. Nattermann and S. Scheidl,
Adv. Phys. {\bf 49}, 607 (2000);T. Giamarchi and S. Bhattacharya cond-mat/0111052 (2001).

\bibitem{giamarchi} T. Giamarchi, and P. Le Doussal, Phys. Rev. B {\bf 55, }6577
(1997); 

\bibitem{grtas96}  D. Ertas, and D. R. Nelson, Physica C {\bf 272}, 79
(1996).

\bibitem{gingras96}  M. J. P. Gingras, and D. A. Huse, Phys. Rev. B {\bf 53}, 15193 (1996).

\bibitem{koshelev98}  A. E. Koshelev, and V. M. Vinokur, Phys. Rev. B {\bf 57}, 8026 (1998).

\bibitem{vinokur98}  V. Vinokur et al.,
Physica C {\bf 295}, 209 (1998).

\bibitem{kierfield98}  J. Kierfield, Physica C {\bf 300}, 171-183 (1998).


\bibitem{mikitik01} G. P. Mikitik, and E. H. Brandt, Phys. Rev. B {\bf 64}, 184514 (2001).

\bibitem{radzyner02} Y. Radzyner, A. Shaulov, Y. Yeshurun, cond-mat 0201024.


\bibitem{pippard69} A. B. Pippard, Philos. Mag. {\bf 19}, 217 (1967).


\bibitem{larkin79} A. I. Larkin and Yu. N. Ovchinnikov, J. Low Temp. Phys. 
{\bf 34}, 409 (1979).

\bibitem{stamopoulos02} D. Stamopoulos and M. Pissas, Phys. Rev. B {\bf 65}, 134524 (2002).

\bibitem{crabtree99}  G. W. Crabtree et al.,
Physics and materials Science of Vortex States, Flux pinning
and Dynamics, 357-385 1999, Kluwer Academic Publishers.

\bibitem{0203337} During the submission of this
paper we became aware of a related work on MgB$_2$ by U. Welp et al.[cond-mat/0203337 (2002)].
The results of this work agree very well with ours.


\bibitem{ling01} X. S. Ling et al.,
\prl {\bf 86},712 (2001).


\end{references}
\end{document}